\documentstyle[aps,preprint,prc]{revtex}

\begin{document}

\begin{center}
{\Large \bf Truncation Method for the Shell Model Calculation\\}
\vspace{.8cm}

{\bf M. Horoi$^{1,2}$, B. A. Brown$^{1}$ and V. Zelevinsky$^{1,3}$\\}

\vspace{.8cm}

{\em
$^{1}$National Superconducting Cyclotron Laboratory,
East Lansing, MI 48824\\

$^{2}$Institute of Atomic Physics, Bucharest Romania\\

$^{3}$Budker Institute of Nuclear Physics, Novosibirsk 630090, Russia\\}

\vspace{42pt}

\end{center}

\vspace{.3cm}

\begin{abstract}

A method of truncating the large shell model basis is
outlined. It relies on the order given by the unperturbed energies of the basis
states and on the constancy of their spreading widths. Both quantities can be
calculated by a simple averaging procedure. The method is  tested in the
sd shell where the JT dimensions are of the order of a few thousand. It proves
to be very effective in the middle of the fp shell where JT dimensions of the
order of a few million are truncated to a few thousand.
\vspace*{.7cm}


\end{abstract}

\newpage

Shell model calculations of the ground state and  low-lying
excited states
are of  interest for our understanding of  nuclear
dynamics and  the (effective) nuclear forces.
 They are also of great interest for the
prediction
and
analysis of various processes
(Gamow-Teller rates, parity nonconservation
matrix elements, electromagnetic transition probabilities,
isospin breaking matrix elements, spectroscopic factors, etc.) important for
nuclear astrophysics and tests of the fundamental interactions in nuclei.
Unfortunately, even for light nuclei, the large
model-space dimensions present a  challenge for the traditional
diagonalization methods (see e.g. Ref. \cite{Wild}).
 During the last few years other approaches
to this problem have been vigorously investigated  \cite{KOf,Feng}.

In this
work we outline a quantitative
method of truncating the  nuclear shell model
spaces to manageable sizes.
To achieve this we show that the basis states,
whose unperturbed energies (the diagonal matrix element of the hamiltonian)
are far away from the lowest one, give relatively  small contributions to the
structure of  the ground states and  low-lying excited states. This statement
can be  quantified due to an interesting property of
the  squared  amplitudes of the basis states (denoted by the
index k),
$\mid C^{\alpha}_{k} \mid^{2}$, as a function of the eigenvalues, E$_{\alpha}$.
Figure 1 presents two of these distributions for the basis state number 2
(left)
and for the basis state number 825 (right). The basis states are ordered by
the JT dimension of the partition (distribution of particles in the
single particle levels) to which they belong. The largest
 partitions are those on
 the left side of k-axis in Fig.~2 (the
dimension of partitions is proportional with
the length of the horizontal thick lines in this figure) while the smallest
partitions are those on the right side of the k-axis. Basis state number
2 belongs to the largest partition, which is situated in the middle of the
unperturbed spectrum ($\bar{E}_{k}$ in Fig.~2), while the
basis state number 825
belongs to a small partition whose unperturbed energy is closer to the
lowest eigenenergy. We note (see left side of Fig.~1)
that the contribution of the basis state number 2 to the ground state
($E_{1} = -135.9$ MeV) is small.
One observes that the distributions of $\mid C^{\alpha}_{k} \mid^{2}$
are close to a Gaussian. It is straightforward to show that their
mean values (centroids) are given by the diagonal matrix elements of the
hamiltonian

\begin{equation}
\bar{E}_{k} \equiv \sum _{\alpha} \mid C^{\alpha}_{k} \mid^{2} E_{\alpha} =
H_{k,k}
\end{equation}

\noindent
and the  widths are  given by

\begin{equation}
\sigma_{k} \equiv \sqrt{\sum _{\alpha} \mid C^{\alpha}_{k} \mid^{2}
E_{\alpha}^{2} - \bar{E}_{k}^{2}} = \sqrt{\sum_{k' \neq k} H_{k', k}^{2}}
\end{equation}

\noindent
The small squares in Fig. 2 represent these quantities for the case of 12
particles in the sd shell with $J^{\pi}T = 0^{+}0$.
The Wildenthal interaction \cite{Wild} has been used for this
plot. We have also used different interactions and  have carried out
 calculations in
the fp
shell, but the results are qualitatively the same. To obtain the points, no
diagonalization is necessary but only a knowledge of the hamiltonian matrix
as given in Eqs. (1) and (2).
Due to the empirical fact that $\sigma$ is nearly
constant for all basis states we can,
for example,
consider only those basis states whose centroids are lower than
$H_{cutoff}=(\bar{E}_{k})_{min} + 3\bar{\sigma}$, where $\bar{\sigma}$ is an
average value for $\sigma_{k}$.

One would like also to avoid the construction of the large hamiltonian matrix.
A useful procedure is to use some average values for the
quantities in Eqs. (1)-(2). One simple way to proceed is to use the
m-scheme average
values given by French and Ratcliff \cite{FR}.
 They are presented in Fig. 2 by the
big squares. They are constant within every partition.
The average values slightly overestimate the exact values
due to the fact that they are derived for the m-scheme, whereas the physical
states
we work with are projected onto good angular momentum and isospin.
However, the m-scheme estimate is good enough for our purpose.

Our method consists of retaining only those partitions whose average
centroids (calculated with the approximate formulae \cite{FR}) are smaller than
$H_{cutoff}$. The method has the advantage that one can
include step-by-step new partitions  in the order of their centroids.
We have tested the method in the sd shell where we know the exact results.
In Fig. 3 we show the results for the lowest $0^{+}1$ states in the case of
10 particles in the sd shell
($\bar{\sigma} = 9.7$ MeV,
($\bar{E}_{k})_{min} + 3\bar{\sigma} = -66.5$ MeV which
corresponds to a  dimension of 310 in Fig. 3).
The left part of Fig.~3 shows the eigenvalues
of the ground state and first excited
state as a function of dimension of the truncated space. The
dimension of the full space is 1132. The filled circles (to the right)
are the exact values. One can see that with  relatively  truncated
spaces one can approach the exact eigenvalues.
To have another measure of
 the precision of the method, we plot in the right part of Fig. 3
the overlaps of the approximate ground state wave function with the exact
ground state wave function (full line).
One can see that with less than 30\% of the full dimension one can
obtain more than 90\% overlaps. The dashed line represents the "optimal"
truncation in the sense that we retained only those basis states whose exact
amplitudes are the highest. One can see that our simple truncation
procedure is near to the "optimal" one.

Our investigations in the sd shell (maximum JT dimension of the order of
a few thousand) show that by  using this method one can reduce the dimension
of the hamiltonian matrices by
typically a factor of 3. Going to  larger model
spaces one might  expect this factor to be even higher,
particularly in cases where
simple shell model configurations (e.g. those with the
lowest $E_{k}$ in the left side of Fig. 2) represent a reasonable
approximation to the exact ground state wave function.
As an example we investigated the $0^{+}1$
low-lying states of $^{54}$Fe (14 particles in the fp
shell, with the Brown-Richter interaction \cite{BRFP})
 using our truncation method. The  dimensions of the problem make the
traditional calculations
unmanageable, even with the next generation of computers: 2229178 JT
dimension and 345400274 m-scheme dimension. The results of our truncation
method are presented in Fig.~4
($\bar{\sigma} = 8$ MeV,
($\bar{E}_{k})_{min} + 3\bar{\sigma} = -143.6$ MeV).
 We don't have exact results with which to
compare, but we can refer to the
result of a recent Monte Carlo calculation \cite{KOFE}
indicated  by the filled circles with errors bars in Fig. 4.
The comparison is encouraging,
and our method should allow the calculation of  energies to
within less than one MeV accuracy for
the ground states and low-lying excited states for even larger model spaces.

Usually, the shell model spaces are truncated according to some qualitative
scheme by retaining only the lowest Hartree-Fock configuration and some
simple 1p-1h or 2p-2h configurations (see e.g. Ref. \cite{NSO} for a recent
survey of this method in the fp-shell). These methods are useful for
some particular class of problems; they are a straightforward extension of
the TDA approximation. They do not fully take into account the details
of the interaction between the valence particles. Our method is more
general; it selects the most important partitions determined by the
interaction, and it is  suited for a hierarchy of successively better
approximations.

A model which has some similarities
 with our method has been presented in Ref. \cite{KRG} and slightly
refined in Ref. \cite{Hasp}. The crucial differences between our approach and
that given in Ref. \cite{Hasp} are: i.) We have proposed a general quantitative
criteria for selecting the most relevant configurations based on the $\sigma$
widths given by the off-diagonal matrix elements of the hamiltonian. The
criteria used in Ref. \cite{Hasp} were based on the observation that one
can use the lowest 33\% of the total configuration as a reasonable
approximation for model spaces with JT dimensions of the order 1000-2000.
Our findings coincide with those from Ref. \cite{Hasp} for similar
dimensions but diverge in some cases
 for larger dimensions. ii.) The truncation scheme
of Ref. \cite{Hasp} was useful only to reduce the diagonalization process;
the JT basis states are still necessary. Our truncation scheme is
able to select the most relevant partitions before
the basis state construction, thus avoiding the calculation of the
full hamiltonian matrix (which is the most time consuming part for any shell
model calculation).

Our method works well in a full major harmonic oscillator
 shell (0$\hbar \omega$ calculations),
where the spurious center-of-mass motion factors out. The standard
method of removing the center-of-mass spurious components of the shell model
wave functions, when many n$\hbar \omega$ excitations are included, is to add
to the nuclear hamiltonian, $H_{N}$, the center-of-mass hamiltonian, $H_{CM}$,
multiplied by a large constant \cite{CM}. In this case the matrix elements of
$H_{CM}$, which dominates,   is different from that of the nuclear
part, and the method described above cannot
be directly applied. One way to circumvent this difficulty is to use our
method to select the most important partitions by taking into account only
the nuclear hamiltonian, $H_{N}$. The part proportional to $H_{CM}$ is
included only during the diagonalization in the truncated basis. A
calculation in $^{14}$N [(0+2+4)$\hbar \omega$], using this scheme reproduces
the energy of the $0^{+}1$ yrast state within 400 keV with a JT dimension
of the truncated space which is one-fifth of the dimension of the full
space (19,498). Work in this direction is in progress.

We also note that the present method might be used as a criteria
for an importance- sampling mechanism for any kind of Monte Carlo method,
such as the "stochastic diagonalization" \cite{STDI}, due to its
 ability to identify
the most important configurations contributing to the structure of the
low-lying state.
It may also be applied to
more general
calculations  such as those  for atoms, molecules or atomic clusters.

\vspace{2cm}

The authors would like to acknowledge support from the
NSF grant 94-03666.

\newpage
\vspace{3cm}

\begin{center}
{\bf Figure captions}
\end{center}

\vspace{1.5cm}

{\bf Figure 1} \ $\mid C^{\alpha} \mid^{2}$ vs $E_{\alpha}$ for basis state
number 2 (left) and basis state number 825 (right).\\

{\bf Figure 2} \
Energy centroids and $\sigma$ widths of the basis states
coefficients distribution (see Eqs. (1)-(2))\\

{\bf Figure 3} \
Energies of the two low-lying states versus the dimension
of the truncated space (left); overlaps of the truncated space
wave functions with the exact ones (right).\\

{\bf Figure 4} \
Energies of the first two low lying states of $^{54}$Fe as
a function of truncation diagonal matrix element (left) and versus the
dimension of the truncated space (right). The filled circles with the
error bars are the result of a recent Monte Carlo calculation \cite{KOFE}.

\end{document}